\input{aipcheck.tex}

\documentclass[
    ,final            
  ]
  {aipproc}

\layoutstyle{6x9}

\def\Im{\mathrm {Im}}

\def\degree{\kern-.2em\r{}\kern-.3em}

\begin{document}

\title{Absolute polarimetry at RHIC}



\author{H.~Okada
\hspace{-2.5mm}
\footnote{Present address: Brookhaven National Laboratory, Upton, NY 11973, USA}
$^{,}\hspace{-2mm}$
}
{
address={Kyoto University, Sakyo-ku, Kyoto 606-8502, Japan}
}
\author{I.~Alekseev}{
 address={Institute for Theoretical and Experimental Physics (ITEP), 117259 Moscow, Russia}
}
\author{A.~Bravar
\hspace{-2.5mm}
\footnote{Present address: University of Geneva , 1205 Geneva, Switzerland}
$^{,}\hspace{-2mm}$
}
{
address={Brookhaven National Laboratory, Upton, NY 11973, USA}
}
\author{G.~Bunce}{
address={Brookhaven National Laboratory, Upton, NY 11973, USA}
,altaddress={RIKEN BNL Research Center, Upton, NY 11973, USA} 
}
\author{S.~Dhawan}{
address={Yale University, New Haven, CT 06520, USA}
}
\author{K.O.~Eyser
\hspace{-2.5mm}
\footnote{Present address: Deutsches Elektronen Synchrotron, 22607 Hamburg, Germany}
$^{,}\hspace{-1.5mm}$
}
{
address={University of California, Riverside, CA 92521, USA}
}
\author{R.~Gill}{
address={Brookhaven National Laboratory, Upton, NY 11973, USA}
}
\author{W.~Haeberli}{
address={University of Wisconsin, Madison, WI 53706, USA}
}
\author{H.~Huang}{
address={Brookhaven National Laboratory, Upton, NY 11973, USA}
}
\author{O.~Jinnouchi
\hspace{-2.5mm}
\footnote{Present address: KEK, Tukuba, Japan}
$^{,}\hspace{-2mm}$
}
{
address={RIKEN BNL Research Center, Upton, NY 11973, USA}
}
\author{Y.~Makdisi}{
 address={Brookhaven National Laboratory, Upton, NY 11973, USA}
}
\author{I.~Nakagawa}{
 address={RIKEN, Wako, JAPAN }
}
\author{A.~Nass
\hspace{-2.5mm}
\footnote{Present address: University of Erlangen, 91058 Erlangen, Germany}
$^{,}\hspace{-2mm}$
}
{
address={Brookhaven National Laboratory, Upton, NY 11973, USA}
}
\author{N.~Saito
\hspace{-2.5mm}
\footnote{Present address: KEK, Tukuba, Japan}
$^{,}\hspace{-2mm}$
}
{
address={Kyoto University, Sakyo-ku, Kyoto 606-8502, Japan}
}
\author{E.~Stephenson}{
address={Indiana University Cyclotron Facility, Bloomington, IN 47408, USA}
}
\author{D.~Sviridia}{
address={Institute for Theoretical and Experimental Physics (ITEP), 117259 Moscow, Russia}
}
\author{T.~Wise}{
address={University of Wisconsin, Madison, WI 53706, USA}
}
\author{J.~Wood
}
{
address={Brookhaven National Laboratory, Upton, NY 11973, USA}
}
\author{A.~Zelenski}{
 address={Brookhaven National Laboratory, Upton, NY 11973, USA}
}

\vspace{-5mm}
\begin{abstract}
Precise and absolute beam polarization measurements are critical for the RHIC spin physics program. 
Because all experimental spin-dependent results are normalized by beam polarization, 
the normalization uncertainty contributes directly to final physics uncertainties.
We aimed to perform the beam polarization measurement to an accuracy of $\Delta P_{beam}/P_{beam} < 5\%$.

The absolute polarimeter consists of Polarized Atomic Hydrogen Gas Jet Target
and left-right pairs of silicon strip detectors and was installed in the RHIC-ring in $2004$.
This system features \textit{proton-proton} elastic scattering in the Coulomb nuclear interference (CNI) region.
Precise measurements of the analyzing power $A_N$ of this process has allowed us to achieve $\Delta P_{beam}/P_{beam}  =4.2\%$ in $2005$ for the first long spin-physics run.

In this report, we describe the entire set up and performance of the system.
The procedure of beam polarization measurement and analysis results from $2004-2005$ are described.
Physics topics of $A_N$ in the CNI region (four-momentum transfer squared $0.001 < -t < 0.032~({\rm GeV}/c)^2$) are also discussed.
We point out the current issues and expected optimum accuracy in $2006$ and the future. 
\end{abstract}
\vspace{-5mm}
\maketitle


\vspace{-10mm}
\subsection{Introduction}
\vspace{-3mm}
	The RHIC spin physics program has been a unique opportunity and important component of the overall RHIC physics program.
	Essential to this spin program are the polarized proton beams to investigate spin-dependent structure in the nucleon.
     Several types of spin-dependent asymmetries in high energy \textit{proton-proton} (\textit{pp}) collisions 
	provide detailed studies of the structure of the proton at a new level of accuracy. 
     Because all experimental results are normalized by beam polarization, $P_{beam}$, the normalization uncertainty contributes directly to final physics uncertainties.
	Therefore accurate and absolute polarization measurements are crucial.
	$P_{beam}$ is obtained from raw asymmetry, $\epsilon_{beam}$, for the transversely polarized proton beam 
     divided by analyzing power, $A_N$, of a certain interaction as shown in Equation~\ref{eq:def1}.
\vspace{-3mm}
	\begin{equation}
		P_{beam}=\frac{\epsilon_{beam}}{A_N}
		\label{eq:def1}
	\end{equation}
	We aimed to achieve an accuracy of $\Delta P_{beam}/P_{beam} < 5\%$ at any beam energy from injection ($24$ GeV/$c$) 
	to flat-top ($100$ GeV/$c$ and $250$ GeV/$c$ in the near future).
	Ideal interactions for polarimetry should satisfy the following conditions:
\vspace{-2mm}
	\begin{enumerate}
		\item well-known or measureable and non-zero analyzing power,
		\item high event rate interaction (large cross-section and/or thicker target) to save data taking time,
		\item similar kinematics for different beam momenta for common detector set up.
	\end{enumerate}
\vspace{-2mm}
	The elastic scattering of the polarized proton beam off a nuclear target A (\textit{p}$^{\uparrow} A \rightarrow$ \textit{p}$A$) 
	in the Coulomb nuclear interference (CNI) region is an ideal process. We choose proton and carbon for A. 
	$A_N$ is a function of four-momentum transfer squared, $-t$. We are looking at very small $-t$ in the order of $10^{-3}$ 
	(GeV/$c$)$^{-2}$.
	We have two types of polarimeters to meet above requirements. One is the RHIC \textit{p}C-polarimeter, which satisfies item $2$ and $3$.
	This polarimeter serves as a semi-on-line beam polarization monitor during the RHIC-run period to tune up the beam acceleration. 
        The RHIC \textit{p}C polarimeter also provides fill-by-fill offline $P_{beam}$ results to experimental groups. 
	However, its accuracy is limited ($\Delta P_{beam}/P_{beam} > 20\%$) mainly due to a difficulty of $-t$ range measurement in \textit{p}C elastic scattering. 
        This difficulty is connected to the need to caliblate each year. 
	The other polarimeter, the {\bf Polarized Atomic Hydrogen Jet Target Polarimeter} (H-Jet polarimeter in short) serves as an absolute calibration of the RHIC \textit{p}C polarimeter. The H-Jet polarimeter satisfies item $1$ and $3$.
	In this report, we focus on the H-jet polarimeter. Details of the RHIC \textit{p}C polarimeter are discussed in~\cite{Itaru}.
	
	The \textit{pp} elastic scattering process is 2-body exclusive scattering with identical particles. 
	$A_N$ for the target polarization and the beam polarization should be same as shown in Equation~\ref{eq:beam0}.
\vspace{-3mm}
	\begin{equation}
		A_N = - \frac{\epsilon_{target}}{P_{target}}=\frac{\epsilon_{beam}}{P_{beam}}, \label{eq:beam0}
	\end{equation}
	$\epsilon_{target}$ is \textit{raw} asymmetry for the \textit{pp} elastic scattering for the
     transversely polarized proton {\bf target} and $P_{target}$ is a well calibrated polarized proton target, which we will discuss later.
	Therefore we can change the role of which is polarized between the target proton and the beam proton. 
	Then the beam polarization is measured as:
\vspace{-3mm}
	\begin{equation}
		P_{beam} =  -P_{target} \frac{\epsilon_{beam}}{\epsilon_{target}}. 
		\label{eq:beam1}
	\end{equation}
	The beauty of the H-Jet polarimeter is that we can cancel the common factors of systematic uncertainty of $\epsilon_{target}$ and $\epsilon_{beam}$.
	By accumulating enough statistics, $\Delta P_{beam}/P_{beam} \approx \Delta P_{target}/P_{target}$ 
	is realizable.
	Although $A_N$ does not appear explicitly in Equation~\ref{eq:beam1}, 
        precise measurements of $A_N$ are very important 
	to confirm that the H-Jet polarimeter works properly at any time. 

	In addition to polarimetry, precise measurements of $A_N$ in the CNI region are important 
        to understand the reaction mechanism completely.
	The \textit{pp} elastic scattering is described in spin-flip and non-flip transition amplitudes. 
	Each amplitude is a sum of the electro magnetic and hadronic forces as functions of $\sqrt{s}$ and $-t$. 
	$A_N$ is expressed as,
\vspace{-3mm}
	\begin{equation}
	  A_N \approx - \frac{\Im[ \phi_{SF}^{em}(s,t)\phi_{NF}^{had*}(s,t)+\phi_{SF}^{had}(s,t)\phi_{NF}^{em}(s,t)]}
                            {|\phi_{NF}^{em}(s,t)+\phi_{NF}^{had}(s,t)|^2}.
		\label{eq:AN_helicity} 
	\end{equation}
	The electro magnetic part of amplitudes ($\phi_{NF}^{em}(s,t)$ and $\phi_{SF}^{em}(s,t)$) are precisely understood by quantum electrodynamics (QED). 
	The hadronic part of the non-flip amplitude ($\phi_{NF}^{had}(s,t)$) , which is related to the unpolarized differential cross-section and total cross-section via the optical theorem at $-t$ =~0, is also understood very well.
	The first term of Equation~\ref{eq:AN_helicity} is calculable 
	 and has a peak around $-t \simeq 0.003 ~({\rm GeV}/c)^2$~\cite{But99} which is generated by proton's anomalous magnetic moment. 

	However, the second term, which includes $\phi_{SF}^{had}(s,t)$, is not well-known. 
	The hadronic reaction in the CNI region is described by non-perturbative quantum chromodynamics (QCD) 
	and a precise prediction is not available. 
	The presence of $\phi_{SF}^{had}(s,t)$ should introduce a deviation in magnitude from the first term and, consequently, there is no precise prediction of $A_N$.
	An initial measurement of $A_N$ in the CNI region was performed by the E$704$ experiment at $200$~GeV/$c$~\cite{E704}. However, precision of data was insufficient for polarimetry.
    
	In the following sections, we will introduce the H-Jet target system, experimental set up and analysis procedures and then we will report on $A_N$ results from RUN$4$.
	We will also report on $P_{beam}$ from RUN$5$ which was the first long spin-physics run.
	Finally, we will discuss current issues and expected optimum precision in $2006$ and the future.

\vspace{-5mm}
\subsection{H-Jet-target system}
\begin{figure}[htp]
	\includegraphics[width=0.6\linewidth, angle=-90]{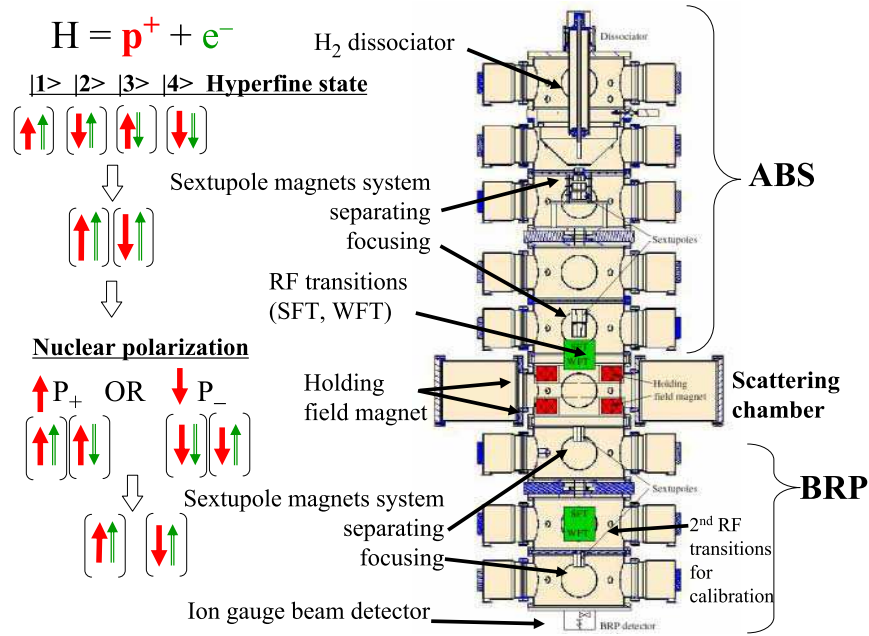}
\vspace{-5mm}
	\caption{H-Jet system overview.}
	\label{fig:H-JetSchematic}
	\end{figure}

\begin{figure}[htp]
	  \includegraphics[width=0.4\linewidth]{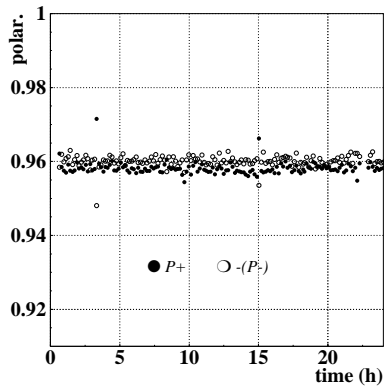}
\vspace{-5mm}
	  \caption{Nuclear polarization measured by BRP in $2004$.}
	  \label{fig:jetrun_bw}
	\end{figure}

\vspace{-3mm}
	The system was installed in the RHIC-ring tunnel for the first time 
        in March $2004$. 
        The commissioning was successfully done.
	Assembly sequence of the system had been completed within $15$ months~\cite{velosity,Yousef}.
	The H-Jet-target system is $3.5$~m in height and approximately $3000$~kg in weight. 
	The target is a free atomic beam, comes from the top in Figure~\ref{fig:H-JetSchematic}, and crosses the RHIC proton beams perpendicularly.
	In this report, we define the negative y-axis as the atomic beam direction and the positive z-axis 
        is the RHIC proton beam direction.
	The velocity of the atomic beam is $1560\pm 20$~m/s~\cite{velosity} and negligible with respect to the RHIC beam.
	The H-Jet-target system is placed on rails along the x-axis.   
	The entire system can be moved along the x-axis by $\pm 10$~mm, in order to adjust 
        the target center to the RHIC beam center.
	As Figure \ref{fig:H-JetSchematic} displays, 
	the system consists of  mainly $3$ parts including nine vacuum chambers and nine differential vacuum stages: 
\vspace{-2mm}
	\begin{enumerate}
		\item Atomic Beam Source, ABS: $1$st to $5$th chambers. Polarize the atomic hydrogen.  
		\item Scattering chamber: $6$th chamber. Collisions between the target-proton 
                      and the beam-proton occur here. 
                      The recoil spectrometers are mounted on both sides of flanges.
		\item Breit-Rabi Polarimeter, BRP: $7$th to $9$th chamber. Measure nuclear polarization, $P_{\pm}$.
	\end{enumerate}
\vspace{-2mm}
	The polarization cycle was ($+/0/-$)=($500/50/500$) seconds and Figure~\ref{fig:jetrun_bw} displays a sample of the measured $P_{\pm}$ in the $2004$ commissioning run~\cite{Zel05}.
	H-Jet-target system was stable over the experimental period.
        The  mean values for nuclear polarization of the atoms:$|P_{\pm}|=0.958 \pm 0.001$.
\vspace{-3mm}	
\vspace{-1mm}	

	The BRP measures the atomic hydrogen polarization, therefore we need to account for 
        the effect on the polarization from background hydrogen molecules. 
	Actually, there were still some molecular hydrogen in the scattering chamber and 
        the measurement was H$_{2}$/H $\sim 0.015$~\cite{velosity}.
	This means that the dilution is about $3\%$ in terms of hydrogen atoms.
	Assuming the molecular hydrogen is unpolarized, the effective target polarization in the $2004$ commissioning run was 
        $P_{target} = 0.924\pm 0.018$.

\vspace{-1mm}
	\begin{figure}[htbp]
        \begin{tabular}{cc}
        \begin{minipage}{0.5\hsize}
	   \raisebox{-5mm}{\includegraphics[width=0.7\linewidth]{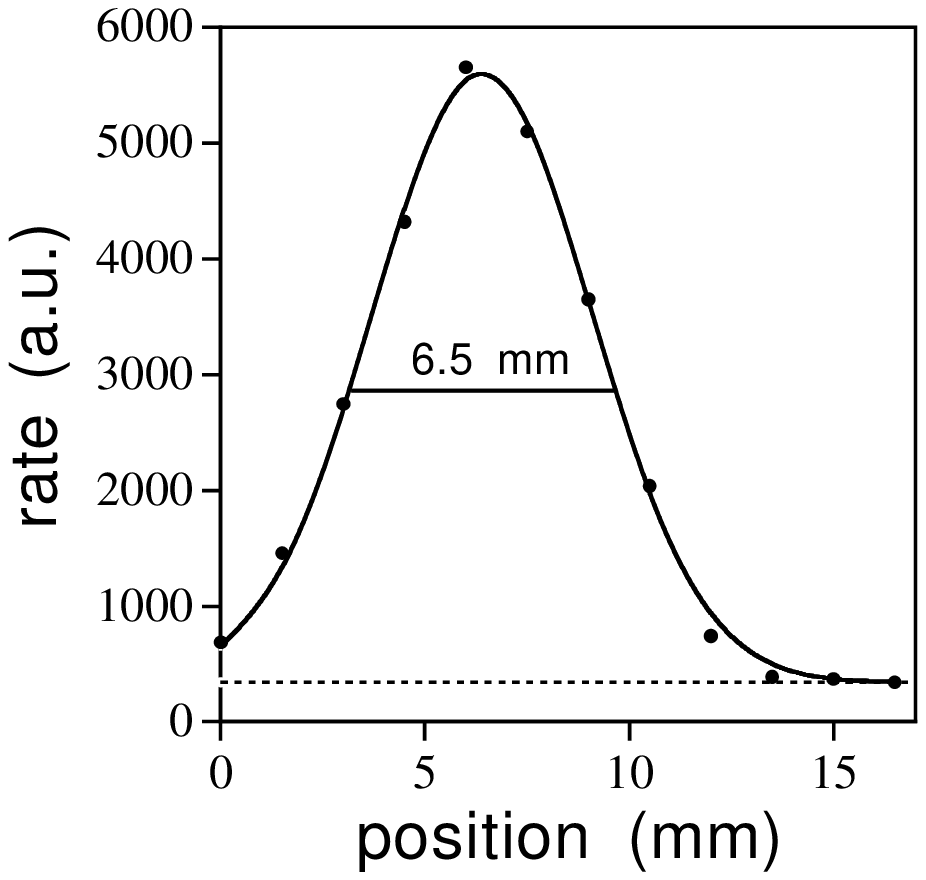}}
\vspace{-2mm}
	   \caption{Atomic beam profile measurements by QMA.}
	   \label{fig:QMA}
	\end{minipage}
        \begin{minipage}{0.5\hsize}
	   \raisebox{5mm}{\includegraphics[width=0.9\linewidth]{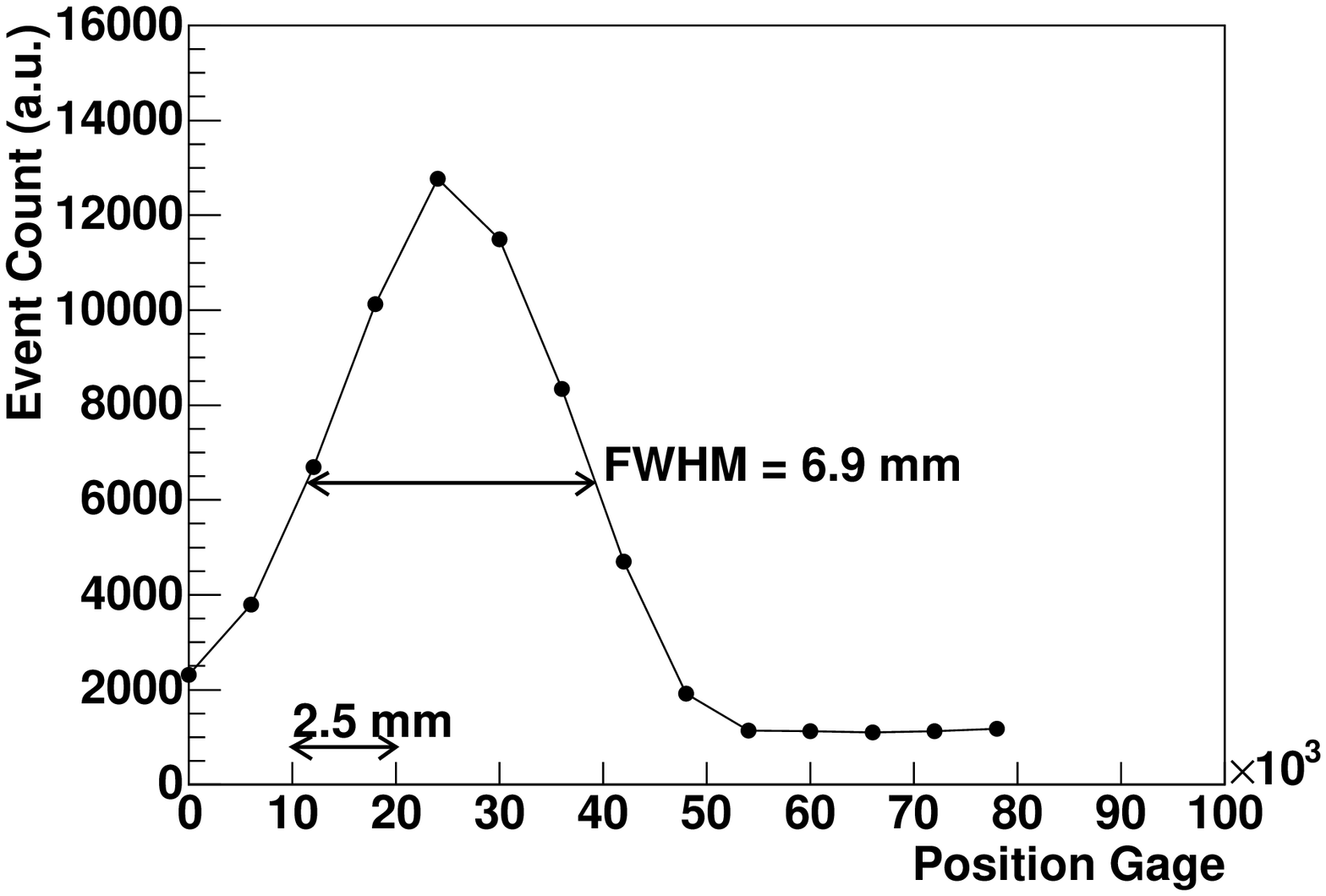}}
\vspace{-2mm}
	   \caption{Left: Atomic beam profile measurements by the compression tube method. 
                    Right: Atomic beam profile measurements by RHIC-beam and the recoil spectrometer.} 
	\label{fig:TargetProfile}
	\end{minipage}
	\end{tabular}
	\end{figure}

	Next, we describe the profile and the density of atomic beam briefly.
	The atomic beam profile was measured with a $2$mm diameter compression tube. 
	The results are displayed in left side of Figure~\ref{fig:TargetProfile}.
	At the center of the scattering chamber, the FWHM of the atomic beam is $6.5$mm and agrees with the design value. 
	Furthermore, we measured the target profile by fixing the RHIC beam ($\sigma \sim 1$~mm) position 
        and moving the entire H-Jet-target system in $1.5$~mm steps. 
        The right side of Figure~\ref{fig:TargetProfile} displays event counts detected by recoil spectrometer 
        versus position.
	Comparing left and right plots of Figure~\ref{fig:TargetProfile}, the two independent measurements agree very well. 
	The target profile measurement using RHIC-beam is important to find the best collision point and estimate the unpolarized background fraction.
	The total atomic beam intensity in the scattering chamber was measured to be $(12.4 \pm 0.2) \cdot 10^{16}$ atoms/s~\cite{Zel05}.
	Taking the measured atomic beam intensity, velocity and profile, the areal target thickness along RHIC beam axis (the z-axis) 
	was calculated to be $(1.3 \pm 0.2)\cdot 10^{12} \rm{atoms/cm}^{2}$~\cite{velosity}.
\vspace{-3mm}	

\subsection{Recoil spectrometer}
\vspace{-3mm}	
The left side of Figure~\ref{fig:layout} displays a schematic layout of the experimental set up to detect \textit{pp} elastic scattering events.
Recoil protons were detected using an array of silicon detectors located to the left and right of the beam 
at a distance $D \simeq 80~{\rm cm}$.
Three pairs of silicon detectors covered an azimuthal angle of $15^\circ$
centered on the horizontal mid-plane.
Detectors were $70.4 \times 50~{\rm mm}^2$ in size, with a $4.4~{\rm mm}$ read out pitch for 
a total of 16 channels per detector. 
We cover recoil protons of kinetic energy of $0.6 \leq T_R \leq 17.0$~MeV.
The recoil angle, $\theta_R$, is obtained by the detector channel number in 
$\simeq~5.5$ mrad steps.
This angular resolution is comparable to the H-Jet-target size.

The silicon detectors were $\sim 400~\mu{\rm m}$ thick.
Recoil protons with kinetic energies,$T_R$, up to 7~MeV are fully absorbed. 
The energy calibration of the silicon detectors was performed using
two $\alpha$ sources $^{241}{\rm Am}$, $5.486$~MeV (and $^{148}{\rm Gd}$, $3.183$~MeV for three out of six detectors). 
Resolution of $T_R$ in the fully absorbed region is $\Delta T_R=0.6$~MeV.
More energetic protons punched through the detectors,
depositing only a fraction of their energy.
Therefore $T_R$ for punch-through protons needs to be corrected 
using the detector thickness and tables for energy loss in silicon~\cite{stoppower}.
The 4-momentum transfer squared is given by $-t = 2 M_p T_R$.
The time-of-flight, TOF, is measured with respect to the bunch crossing
timed by the accelerator RF clock. 
The estimated TOF resolution is $\Delta{\rm TOF} \simeq 3~{\rm nsec}$ and a result of the intrinsic time resolution
of the detectors ($\leq 2~{\rm nsec}$) and the length of the RHIC beam bunches ($\sigma \simeq 1.5~{\rm nsec}$).
Details of the recoil spectrometer and analysis for RUN$4$ are discussed in~\cite{hiromiD}.
\begin{figure}[t]
\begin{tabular}{cc}
\begin{minipage}{0.45\hsize}
  \raisebox{-10mm}{\includegraphics[width=0.7\linewidth,angle=-90]{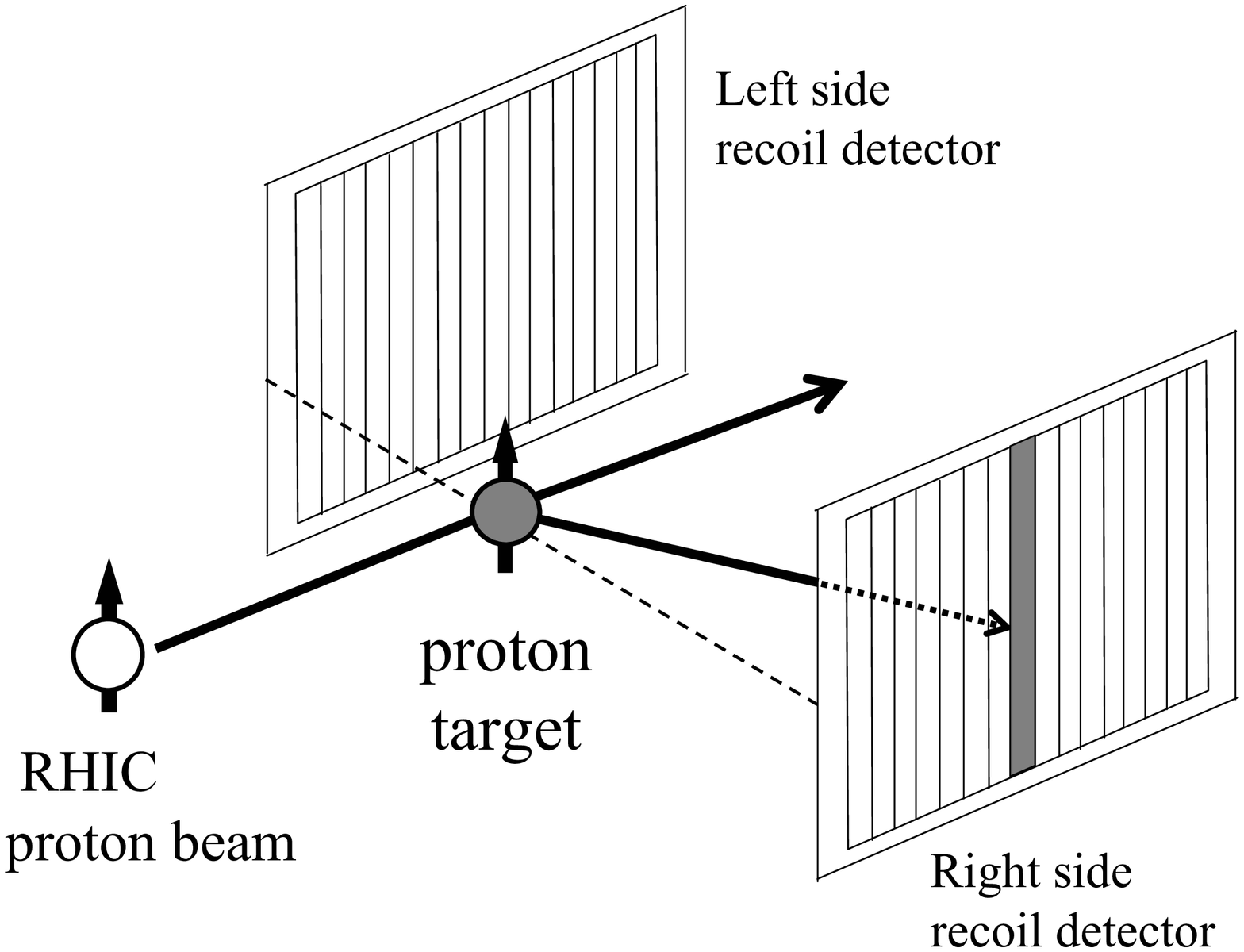}}
\vspace{-2mm}
  \caption{Sketch of left-right pair of silicon detectors.}
  \label{fig:layout}
\end{minipage}
\begin{minipage}{0.55\hsize}
  \raisebox{-10mm}{\includegraphics[width=1.0\linewidth, angle=0]{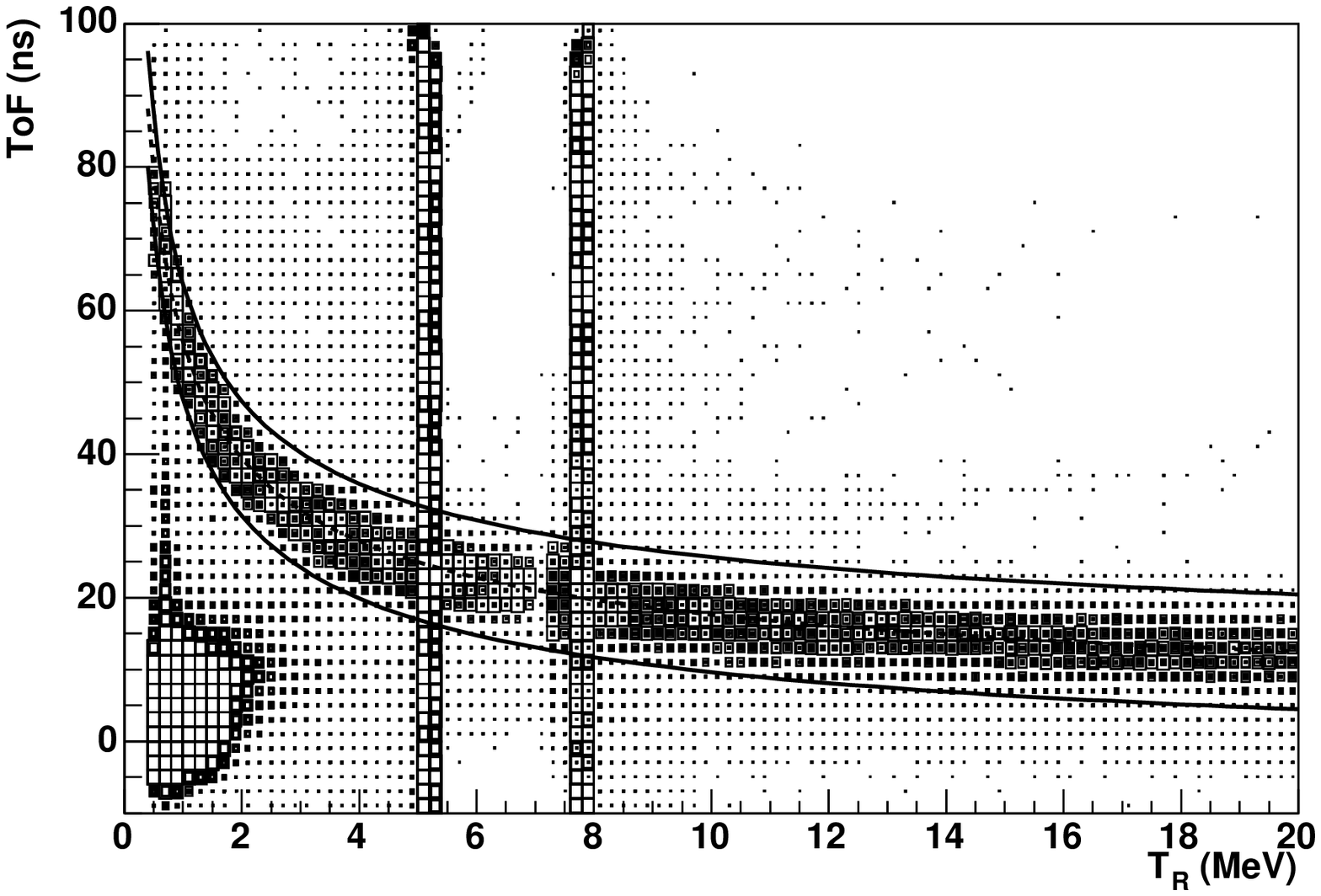}}
\vspace{-2mm}
  \caption{Left:Sketch of left-right pair of silicon detectors. 
     Right: The correlation between $TOF$ and the incident energy, 
     $T_R$, in one of the silicon detectors. 
     Two solid lines corresponds to $\pm 8$~nsec shifted with respect 
     to the expected TOF value for given $T_R$ from Equation~\ref{eq:ToFCalc}.}
  \label{fig:layout}
\end{minipage}
\end{tabular}
\end{figure}

\vspace{-3mm}	
\subsection{Elastic event selection}
\vspace{-3mm}	

In the \textit{pp} elastic scattering process, both the forward-scattered particle 
and the recoil particle are protons and no other particles are produced in the process. 
The elastic process can be 
identified by detecting the recoil particle only, by identifying the recoil particle as a proton throughout the relation of
TOF and $T_R$, and by observing that the missing mass of the forward scattered system is the proton mass.
Recoil protons were identified using the
non-relativistic relation
\vspace{-3mm}	
\begin{equation}
	T_R = \frac{1}{2} M_p (D/{\rm TOF})^2.
	\label{eq:ToFCalc}
\end{equation}

The right plot of Figure~\ref{fig:layout} displays $T_R$ and TOF correlation from one detector for $16$ channels.
We can see recoil protons clearly around the expected TOF value for $T_R$.  
In this figure, the energy for punch-through events have been corrected~\cite{hiromiD,hiromi06}.
The events which are vertically distributed around $5.5$~MeV are from the calibration $\alpha$ source ($^{241}{\rm Am}$).
(The punch-through correction causes another vertical distribution around $7.5$~MeV.) 
Events less than $3$~MeV and less than $30$~nsec are prompt particles, which are possibly pions 
from beam-related interactions upstream.
Events were selected in a TOF interval of $\pm 8$~nsec
around the expected TOF value for recoil protons of a
given $T_R$ as shown in two lines in the figure.

\begin{figure}[t]
  \includegraphics[width=0.4\linewidth, angle=0]{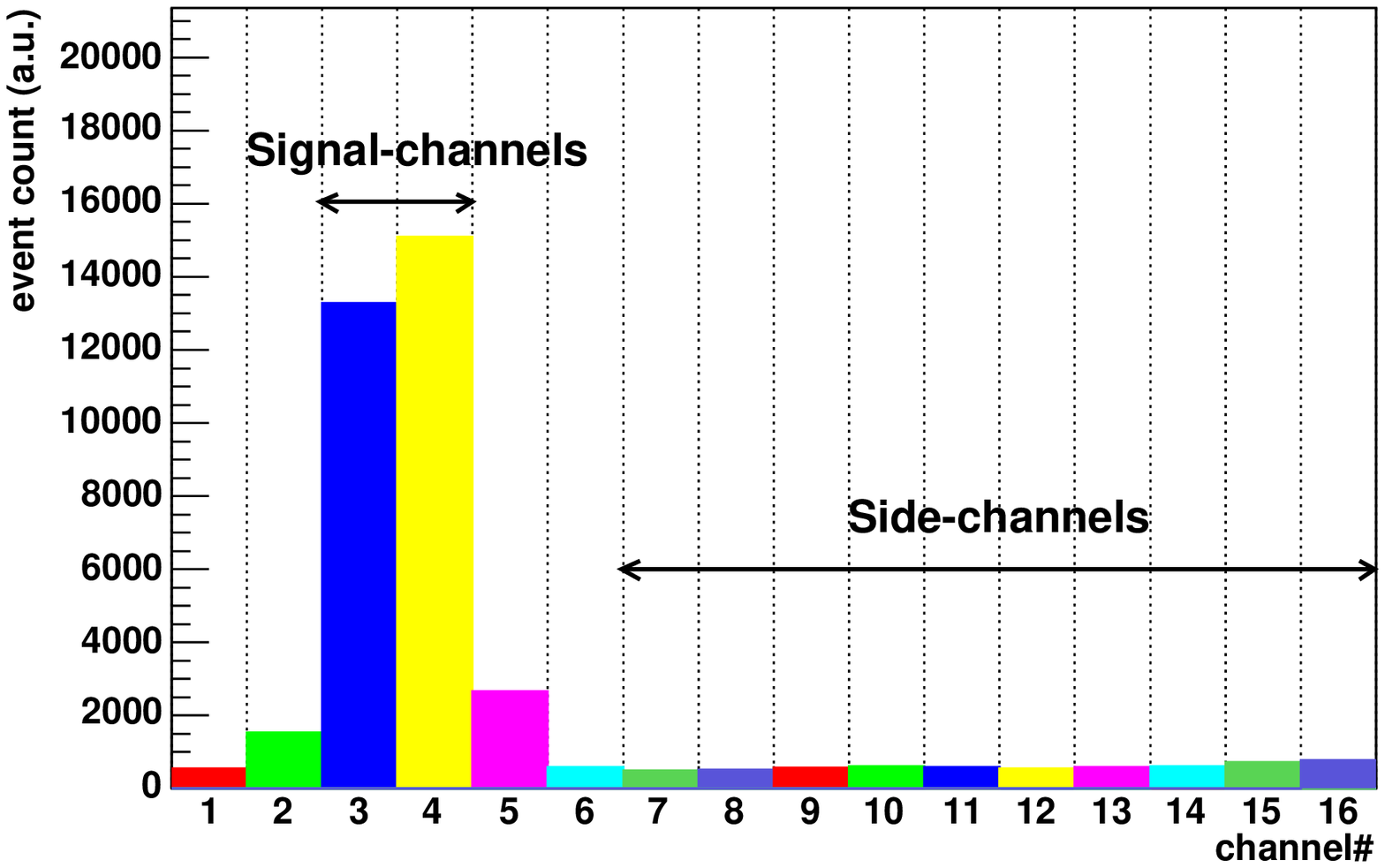}
\vspace{-5mm}
  \caption{Event distribution of a certain $T_R$ interval as function of channel number.}
   \label{fig:Chan}
\end{figure}

On the basis of the measured $\theta_R$ and $T_R$,
the mass of the undetected forward scattered system
(the missing mass $M_X$)
can be reconstructed,
\vspace{-3mm}	
\begin{equation}
 	   {M_X}^2 = {M_p}^2-2\Bigl( (M_p+E_1)T_R - \sin \theta _R \sqrt{2M_pT_R(E_1^2-M_p^2)}\Bigr), \label{eq:mx}
\end{equation}
where $E_1$ is the energy of the incident beam proton.
For $pp$ elastic scattering, events are identified on the basis of the $\theta_R$-$T_R$ relation
\vspace{-3mm}	
\begin{equation} 
	T_R=2M_p \sin^2 \theta_R \frac{E_1-M_p}{E_1+M_p},\label{eq:mp}
\end{equation}
which is obtained applying $M_X = M_p$ in Equation~\ref{eq:mx}.
The difference for $E_1=24$~GeV and $E_1=100$~GeV, the two beam energies reported here, 
is $\sim 3$~mrad at $T_R=17$~MeV and smaller shift at lower energies 
Figure~\ref{fig:Chan} displays the event distribution of a certain $T_R$ interval as function of channel number. 
For each $T_R$ bin $pp$ elastic events were selected in the proper detector strips 
centered around the expected $\theta_R$ angle.

The channel for diffractive dissociation opens at $M_X > M_p + M_{\pi} = 1.08~{\rm GeV}/c^2$.
The kinematical boundary for $M_X=M_p+M_{\pi}$ is given by Equation~\ref{eq:mx} and is out of the acceptance for 
$E_1=24$~GeV.
For $E_1=100$~GeV, the kinematical boundary for the $M_X=M_p+M_{\pi}$ is inside the acceptance for $T_R>8$~MeV.
But the contamination is estimated to be less than 0.5\% from $M_X$ spectra.

The selected event yield is sorted by $T_R$ bins.
We collected $4.3$~M events in fourteen $T_R$ bins at $100$~GeV/$c$ and $0.8$~M events in nine $T_R$ bins at $24$~GeV/$c$ 
in the region $0.001~\leq-t~\leq~0.035$~(GeV/$c$)$^2$ ($0.5~\leq~T_R~\leq~17$~MeV) using the "clock-wise" beam. 
Furthermore, the selected event yield in each $T_R$ bin is sorted by spin states (beam, target, up-down) and the detector side (left-right).  
Finally, we calculate \textit{raw} asymmetries of target or beam polarization using the square-root formula:
\vspace{-2mm}	
\begin{equation}
\epsilon = \frac{\sqrt{N_{\uparrow}^L \cdot N_{\downarrow}^R}-
            \sqrt{N_{\uparrow}^R \cdot N_{\downarrow}^L}}
           {\sqrt{N_{\uparrow}^L \cdot N_{\downarrow}^R}+
            \sqrt{N_{\uparrow}^R \cdot N_{\downarrow}^L}},
	\label{eq:epsilon}
\end{equation}
where if we sort by H-Jet-target (beam) polarization, we have $\epsilon_{target}$ ($\epsilon_{beam}$). 
This expression cancels luminosity and acceptances asymmetries.

\vspace{-3mm}	
\subsection{$A_N$ measurements from RUN$4$}
\vspace{-3mm}	
$A_N$ data are obtained as follows:
\vspace{-3mm}	
\begin{equation}
A_N = - \frac{\epsilon_{target}}{P_T}\frac{1}{1-R_{BG}},
	\label{eq:AN0}
\end{equation}
where $R_{BG}$ is the background levels for each $T_R$ bin.
The backgrounds consisted of (a) $\alpha$ particles from the calibration sources, 
(b) beam scraping, and (c) beam scattering from the unpolarized residual target gas.
The dominant component was (c), due to unfocused molecular hydrogen, and was accounted for as a dilution of the target polarization.
Therefore, $R_{BG}$ is estimated to be $0.02 \sim 0.03$ from (a) and (b)~\cite{hiromi06}. 

\begin{figure}[htbp]
	\includegraphics[width=0.6\linewidth]{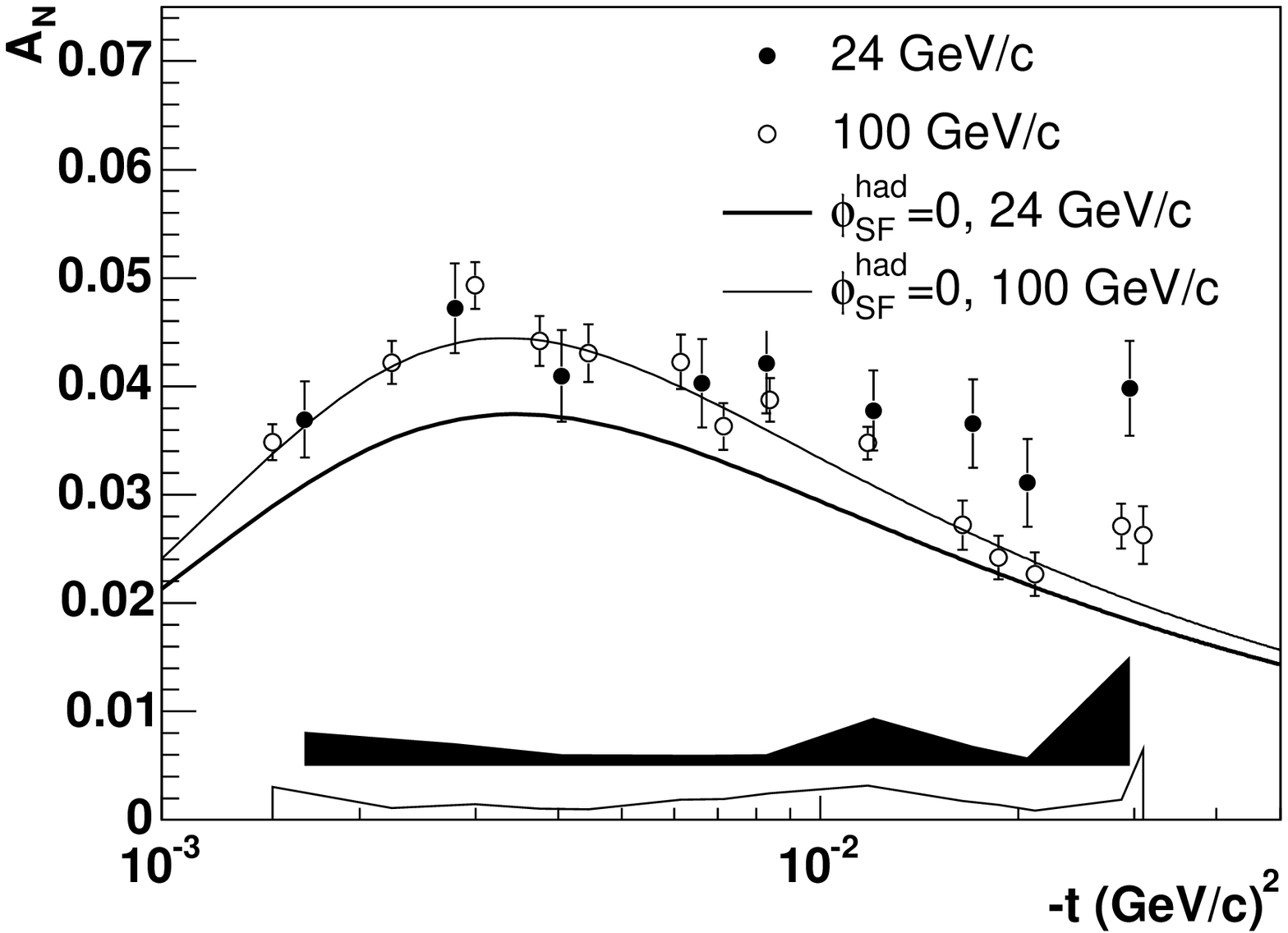}
\vspace{-5mm}
	\caption{$A_N$ data at $24$~GeV/$c$ and $100$~GeV/$c$. 
                 The uncertainties shown are statistical and lower bands 
                  represent the systematic ones.
                  Thick and thin solid lines are the QED prediction 
                 without $\phi_{SF}^{had}$.}
	\label{fig:AN_24_100_PLB}
\end{figure}

Figure~\ref{fig:AN_24_100_PLB} displays the results for $A_N$ at $100$~GeV/$c$ (open circles) and $24$~GeV/$c$ (black filled circles) in $2004$.
The uncertainties shown are statistical. The lower bands represent the total systematic uncertainties.
The thick and thin solid lines are the QED prediction with no spin-dependent hadronic contribution ($\phi_{SF}^{had}$) and 
corresponds to the first term in Equation~\ref{eq:AN_helicity} at these beam momenta.

\vspace{-3mm}
\vspace{-1mm}
Sources of systematic uncertainties come from 
$T_R$ bin-dependent and overall normalization:
(1) the uncertainty on the target polarization giving an overall $\Delta P_{target}/P_{target}=2.0\%$ 
normalization uncertainty; 
(2) the left-right detector acceptance asymmetry;  
(3) event selection criteria; and 
(4) background contribution from (a) and (b).
The major component was (2) at the lowest and highest $T_R$ bins from detector edges. 
We also have a relatively large acceptance asymmetry at the punched-through energy region.

For the $A_N$ measurements, we averaged beam spin up-down states to obtain unpolarized beam.
The difference of absolute value of beam polarization between spin-up and spin-down states was confirmed to be small by acquiring the results from the RHIC-\textit{p}C polarimeter. 
Therefore the residual components beam polarization has no effect on this result.

The $A_N$ data at $24$~GeV/$c$ ($\sqrt{s}=6.8$~GeV) and $100$~GeV/$c$ ($\sqrt{s}=13.7$~GeV) 
are consistent in the region of $-t < 10^{-2}$.
However, these $A_N$ results at different $\sqrt{s}$ energies indicate a $\sqrt{s}$ dependence of $\phi_{SF}^{had}(s,t)$.
The $A_N$ data at $\sqrt{s}=6.8$~GeV are \textit{not} consistent with the solid line 
($\chi^2$/ndf=$35.5/9$) and this discrepancy implies the presence of a hadronic spin-flip contribution, 
$\phi_{SF}^{had}(s,t)$~\cite{hiromi07}.
On the other hand, the $A_N$ data at $\sqrt{s}=13.7$~GeV are consistent 
with the QED prediction ($\chi^2$/ndf=$13.4/14$)~\cite{hiromi06}.  

The theoretical efforts to determine $\phi_{SF}^{had}(s,t)$ including its $\sqrt{s}$ dependence are ongoing.
Using experimental results ($A_N$ in \textit{pp} elastic scattering at $\sqrt{s}=13.7$~GeV and in \textit{p}C elastic scattering at 
$\sqrt{s}=6.4$~GeV~\cite{Toj02} and $13.7$~GeV~\cite{Jin04}) as input parameters, 
prediction for $A_N$ at $\sqrt{s}=6.8$~GeV was given in recent work~\cite{True07}. 
The prediction suggested a significant $\sqrt{s}$ dependence of $\phi_{SF}^{had}(s,t)$, and agreed with our data within $1$-$\sigma$ uncertainty.
More comparisons between theoretical prediction and further experimental data at different beam momenta are required to understand $\phi_{SF}^{had}(s,t)$.

\vspace{-3mm}
\subsection{$P_{beam}$ results from RUN$5$}
\begin{figure}[htp]
	\raisebox{-5mm}{\includegraphics[width=0.65\linewidth]{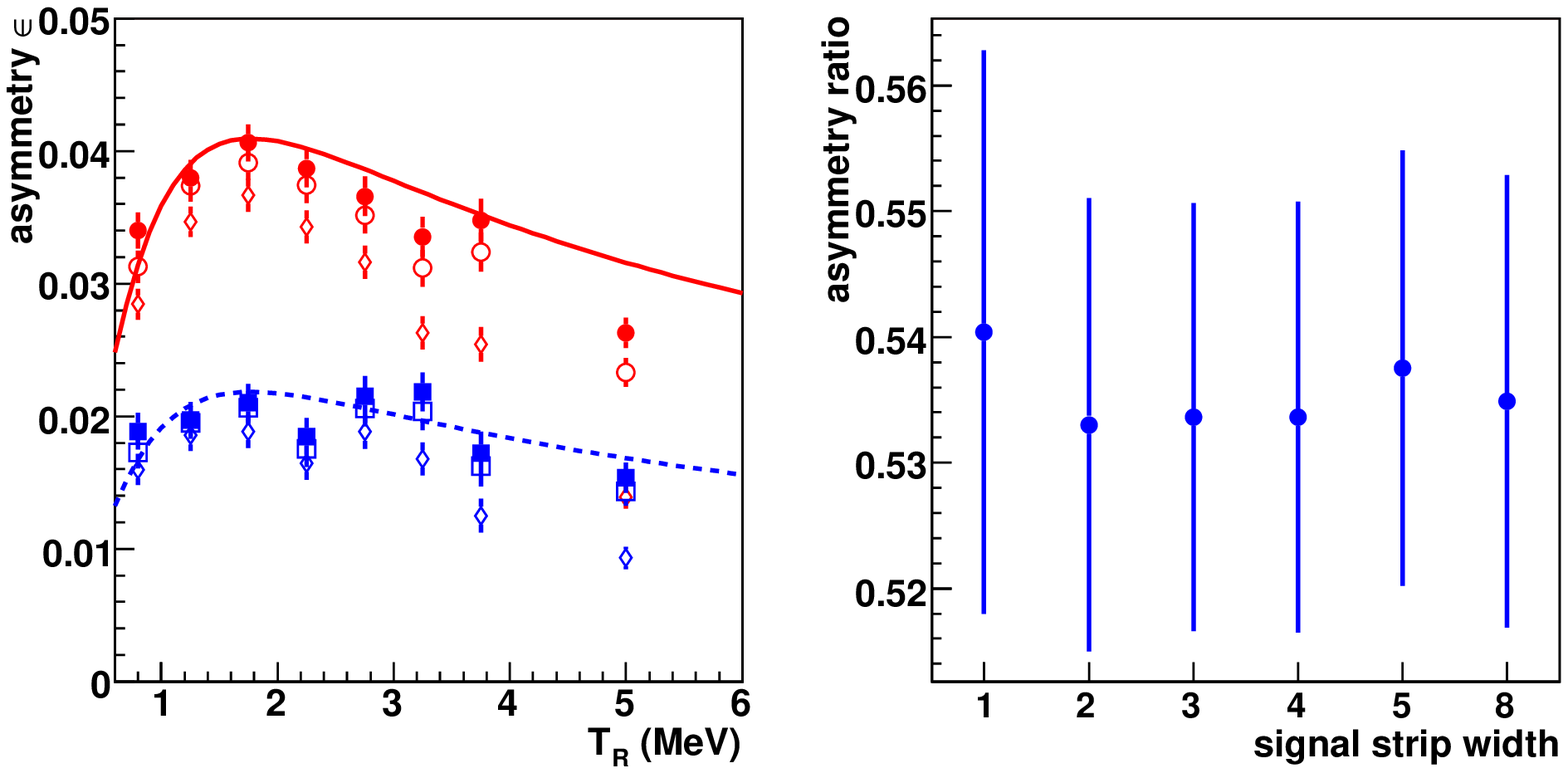}}
	\caption{Background study for the "clock-wise" beam of RUN$5$. 
         Left: Filled circle refers to two strips for the original asymmetry. 
               Open circle and open diamond symbols refer to four and eight strips. 
               Upper group is $\epsilon_{target}$ and lower group is $\epsilon_{beam}$.
         Right: Asymmetry ratio as a function of signal strip width.} 
	\label{fig:RUN5_BLUE}
\end{figure}

\begin{figure}[htp]
	\raisebox{-5mm}{\includegraphics[width=0.65\linewidth]{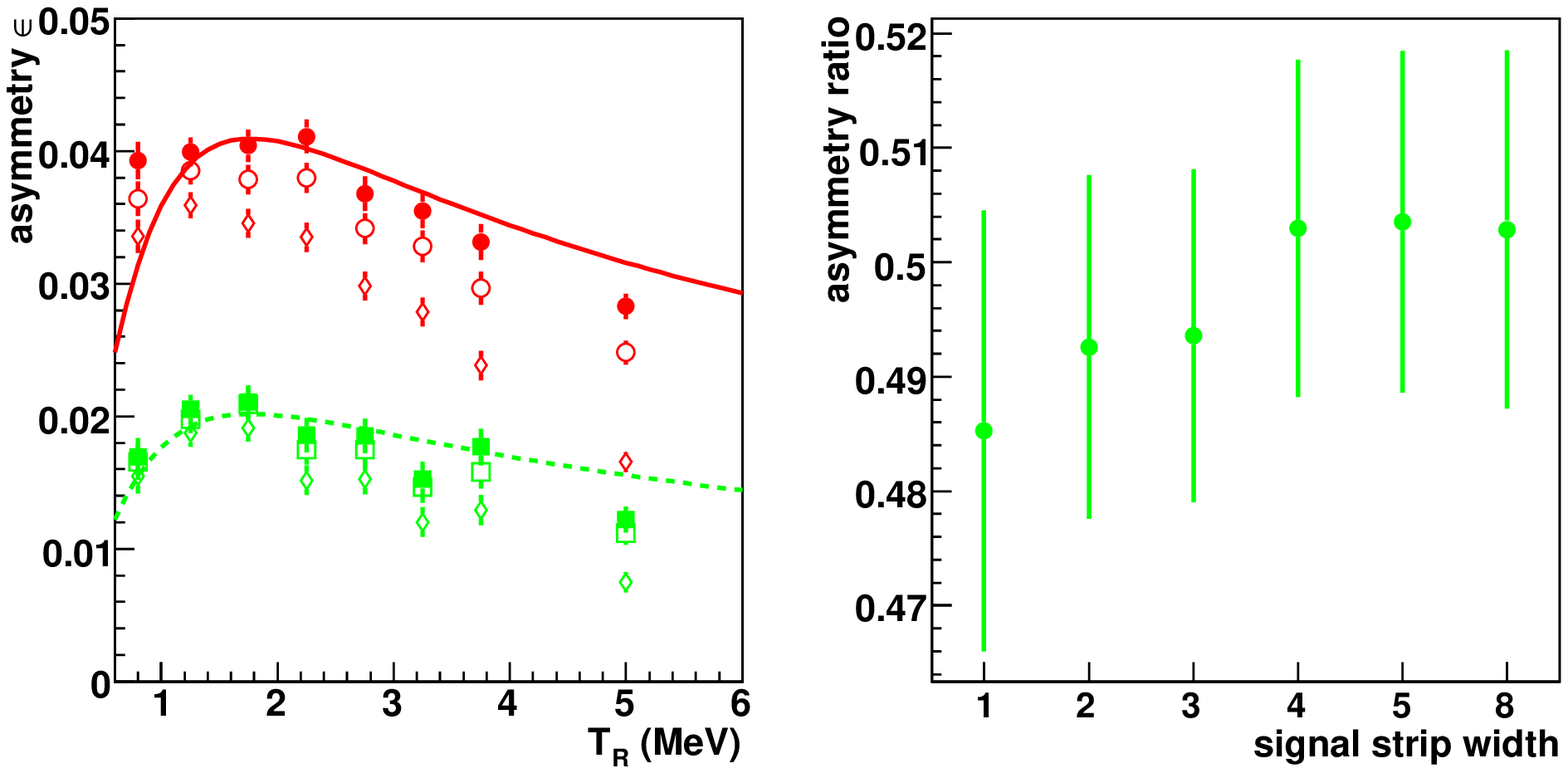}}
	\caption{Same as Figure~\ref{fig:RUN5_BLUE} for the "counter-clock-wise" beam of RUN$5$.}
	\label{fig:RUN5_YELLOW}
\end{figure}

\vspace{-3mm}
In $2005$, one of the two RHIC beams was centered on the H-Jet-target for several days to 
accumulate enough statistics for a precise measurement of the beam polarization.
We displaced the "unused" beam approximately $10$~mm horizontally and vertically from the H-Jet-target center. 
Both beams were measured repeatedly over the course of a few weeks.
Detailed experimental set up and analysis for RUN$5$ are discussed in~\cite{Oleg}. 
We accumulated $5.3$~M events for the "clock-wise" beam and $4.2$~M events for the "counter-clock-wise" beam.
$\epsilon_{target}$ and $\epsilon_{beam}$ for both beams were calculated using Equation~\ref{eq:epsilon}.
We confirmed that $A_N=-\epsilon_{target}/P_{target}$ from both beams were consistent with $A_N$ of $2004$ results. 
For polarimetry use, we use data in the peak asymmetry region of $1 \leq T_R \leq 4$~MeV 
to eliminate acceptance asymmetry and prompt events.
Then, $P_{beam}$ is obtained using Equation~\ref{eq:beam1}.
The total systematic uncertainty in $2005$ was $\Delta P_{target}^{sys.}/P_{target}=2.9\%$.
The dominant two components were: 
\vspace{-2mm}
\begin{itemize}
	\item backgrounds from residual gas and displaced (not used) beam ($2.2\%$),
        \item uncertainty on the target polarization giving an overall $\Delta P_{target}/P_{target}=2.0\%$.
\end{itemize} 
\vspace{-2mm}
Studies of backgrounds were carried out by varying the measured background contributions near the elastic \textit{pp} signal.
The strip distributions show a uniformly spread yield over the non-signal strips. 
(An example at a certain $T_R$ interval is shown in Figure~\ref{fig:Chan}.)
By increasing the number of strips used for the elastic peak, the background contribution can be increased in a controlled way.
Figures~\ref{fig:RUN5_BLUE} and~\ref{fig:RUN5_YELLOW} explain this study of the "clock-wise" beam and the "counter-clock-wise" beam.
The right parts of these figures summarize the asymmetry ratios for different number of strips for the signal region, 
going from one to eight.
The original asymmetries were calculated with two strips.
The open circle and open diamond symbols on the left refer to four and eight strips, thereby doubling and quadrupling the background contributions.
Asymmetry ratio of the "clock-wise" beam seems to drop slightly, and that of the "counter-clock-wise" beam rises, but the variations are smaller than the statistical uncertainties.
Therefore, these differences do not necessarily point to a polarization dependence of inelastic events. 
Also, no clear asymmetry has been seen in background events.
Absolute beam polarizations of the "clock-wise" and the "counter-clock-wise" beams at $100$~GeV/$c$ in $2005$ are $49.3 \% \pm 1.5 \% {\rm (stat.)} \pm 1.4 \% {\rm (sys.)}$ and $44.3 \% \pm 1.3 \% {\rm (stat.)} \pm 1.3 \% {\rm (sys.)}$.
We achieved accurate beam polarization measurement $\Delta P_{beam}/P_{beam} =4.2\%$.

\vspace{-5mm}
\subsection{Future prospect}
\vspace{-3mm}
Finally, the current issues and expected optimum precision in $2006$ and the future are discussed here briefly.
More data are collected in RUN$6$, $8.2$~M events for the "clock-wise" beam and $10.7$~M events for the "counter-clock-wise" beam at $100$~GeV/$c$.
The expected statistical uncertainty is approximately $1$~\%.
More detailed study of background contribution to systematic contribution is ongoing. 
An improvement to reduce the uncertainty for the unpolarized fraction of the H-Jet-target is required for a breakthrough to a new level of accuracy. 

$A_N$ data at different beam energies are an important physics topic.
In RUN$6$, we also took $31$~GeV/$c$ data with better statistics than RUN$4$ $24$~GeV/$c$.
These data will contribute to a comprehensive understanding of $\phi_{SF}^{had}(s,t)$.



\bibliographystyle{aipprocl} 






\end{document}